\documentclass[pre, aps,showpacs,supergroupedaddress,epsfig,amsmath,amssymb]{revtex4}
\usepackage{epsfig,amsmath,amssymb,bm,epsf,graphics,psfrag,verbatim,subfigure}
\usepackage[usenames,dvipsnames]{color}
\usepackage{bbm}
\usepackage{mathrsfs}
\usepackage{amsfonts}
\usepackage{color}
\usepackage{pbox}
\def\newblock{\hskip .11em plus .33em minus .07em}
\def\newsub{{\rm norm}}

\newcommand{\be}{\begin{equation}}
\newcommand{\ee}{\end{equation}}
\newcommand{\ba}{\begin{eqnarray}}
\newcommand{\ea}{\end{eqnarray}}
\newcommand{\bw}{\begin{widetext}}
\newcommand{\ew}{\end{widetext}}
\newcommand{\Tr}{{\rm{Tr}\,}}

\newcommand{\xv}{{\bm{x}}}
\newcommand{\yv}{{\bm{y}}}
\newcommand{\zv}{{\bm{z}}}

\newcommand{\cv}{{\bm{c}}}
\newcommand{\uv}{{\bm{u}}}

\newcommand{\xh}{{\hat{\bm{x}}}}
\newcommand{\yh}{{\hat{\bm{y}}}}
\newcommand{\zh}{{\hat{\bm{z}}}}
\newcommand{\Qm}{{\bm{Q}}}

\begin{document}
\title{Generalized Deam-Edwards Approach to the\\
Statistical Mechanics of Randomly Crosslinked Systems}
\author{Xiangjun Xing$^1$}
\author{Bing-Sui Lu$^1$}
\author{Fangfu Ye$^2$}
\author{Paul M. Goldbart$^2$}
\affiliation{%
$^1$Department of Physics and Institute of Natural Sciences,
Shanghai Jiao Tong University, Shanghai, China}
\affiliation{$^2$School of Physics,
Georgia Institute of Technology,
837 State Street, Atlanta, GA 30332-0430, U.S.A.}

  \date{\today} 
\begin{abstract}
We address the statistical mechanics of randomly and permanently crosslinked networks.
We develop a theoretical framework (vulcanization theory) which can be used to systematically analyze the correlation between the statistical properties of random networks and their histories of formation.
Generalizing the original idea of Deam and Edwards, we consider an instantaneous crosslinking process, where all crosslinkers (modeled as Gaussian springs) are introduced randomly at once in an equilibrium liquid state, referred to as the preparation state.
The probability that two functional sites are crosslinked by a spring exponentially decreases with their distance squared.
After formally averaging over network connectivity, we obtained an effective theory with all degrees of freedom replicated $1+n$ times.
Two thermodynamic ensembles, the preparation ensemble and the measurement ensemble, naturally appear in this theory.
The former describes the thermodynamic fluctuations in the state of preparation, while the latter describes the thermodynamic fluctuations in the state of measurement.
We classify various correlation functions and discuss their physical significances.  In particular, the memory correlation functions characterize how the properties of networks depend on their history of formation, and are the hallmark properties of all randomly crosslinked materials.
We clarify the essential difference between our approach and that of Deam-Edwards, discuss the saddle-point order parameters and its physical significance.
Finally we also discuss the connection between saddle-point approximation of vulcanization theory, and the classical theory of rubber elasticity as well as the neo-classical theory of nematic elastomers.

\end{abstract}

\maketitle

\section{Introduction}
\label{sec:intro}
Randomly crosslinked systems---including rubbers, gels, liquid crystalline elastomers, cytoskeleton networks, and many other related systems---arise widely in nature and are also extensively manufactured by scientists and technologists . Their physical properties are dominated by heterogeneous, system-spanning random networks.  A proper specification of this random network structure therefore constitutes the first step towards obtaining a more complete understanding of these complex materials.

In a typical protocol for making randomly crosslinked materials, one starts from a liquid system {\em under particular  physical conditions} (e.g., of temperature, pressure, solvents, etc.), and randomly introduces crosslinkers that link together nearby particles or polymers.
The macroscopic liquid state prior to crosslinking is defined as  {\em the preparation state}.  A realistic crosslinking process necessarily takes some finite interval of time, and is therefore inevitably accompanied by some irreversible responses of the system.  Inclusion of these irreversible aspects would make the theoretical modeling of the random crosslinking process extremely complicated.
To simplify the analysis and nevertheless capture the essential physics of random crosslinking, we shall therefore follow the original idea of Deam and Edwards~\cite{Deam-Edwards-1976} by considering an idealized, instantaneous crosslinking scheme in which all crosslinkers are introduced into the system at one particular instant in time.
If a sufficient number of crosslinkers is introduced, a system-spanning
random network emerges, and endows the material with a non-vanishing shear modulus.  A continuous transition such as this, from a liquid to a random solid, is usually called {\em gelation}---or {\em vulcanization} in settings of pre-existing polymers.  Gelation and vulcanization are quite different from usual thermodynamic phase transitions, because they are irreversible and the resulting systems acquire intrinsic spatial heterogeneity.  The statistical theory for these transitions is therefore necessarily more complicated.


It is well known that the physical properties of randomly crosslinked materials depend both on the state of measurement and on the state of preparation (i.e, the history of formation).   This is, of course, a feature common to all heterogeneous materials, and has been  widely appreciated by experimentalists.  It was also explicitly pointed out by Deam and Edwards~\cite{Deam-Edwards-1976} and by de Gennes~\cite{polymer:deGennes} as early as the 1970's.  It therefore follows that the statistical physics of randomly crosslinked materials involves two statistical ensembles:
{\em the preparation ensemble} and
{\em the measurement ensemble}.
This necessitates a substantial generalization of the standard Gibbs ensemble theory for equilibrium statistical mechanics.

To substantiate this seemingly extreme statement, let us discuss the remarkably distinct properties exhibited by {\em nematic} elastomers that have been crosslinked under distinct conditions.  Nematic elastomers~\cite{LCE:WT} are crosslinked nematic polymer melts, and their liquid crystalline ordering is strongly coupled to their network elasticity.  Two crosslinking protocols have been studied recently by Urayama {\it et al\/}.
In one, the system is first crosslinked in the high-temperature, isotropical phase of liquid crystallinity, and is then brought to a lower-temperature, nematic phase.  The resulting system (called an isotropic-genesis polydomain (PD) nematic elastomer, or I-PNE) exhibits a polydomain nematic state, with a typical nematic domain size of a few microns.  Upon stretching, the nematic domains gradually rotate and deform, and eventually, beyond certain strain value, they align along the external stress so as to form a monodomain (MD) state.
The stress response during the PD-MD transition window is extremely soft, being orders of magnitude lower than that of typical elastomeric materials.

In the other protocol, the system is first quenched into the nematic phase, so that the nematic order is well developed locally, but is still frustrated by nematic defects at larger scales
(typically beyond $100\,\mu{\rm m}$).
Crosslinking carried out in this state freezes in the patterns of nematic defects.  The resulting system (known as a nematic-genesis polydomain nematic elastomer, or N-PNE) also exhibits a stress-driven PD-MD transition.  The observed stress is, however, within the range of typical elastomer materials, and much higher than that of I-PNEs.
The striking differences between I-PNEs and N-PNEs are entirely due to their distinct preparation states.   Evidently, a theoretical understanding of these differences demands a statistical theory that involves both the preparation ensemble and the measurement ensemble.


A generic protocol for preparing and measuring a randomly crosslinked system has three steps, each statistical in nature.
First, there is a preparation state.  We shall denote averaging over the thermal fluctuations in this state by square brackets
$\left[ \, A \, \right]_{\rm p}$\,, where the subscript p stands for
\lq\lq preparation\rlap.\rq\rq\thinspace\
Second, there is an instantaneous crosslinking stage, at which crosslinkers are randomly introduced into the system.  Averages over realizations of crosslinkers are denoted by an overbar:
$\overline{\, A \, }$ .
Finally, there is a measurement state.  Averages over fluctuations in this state are denoted by angle brackets
$\langle A \rangle_{\rm m}$\,,
where the subscript m stands for
\lq\lq measurement\rlap.\rq\rq\thinspace\
Note that these three types of average have a {\it causal ordering in time\/}:
Any quantity averaged in a measurement state must also be averaged over crosslinker realizations and over the preparation state, as the latter two ensembles constitute the history of formation.  By contrast, quantities averaged in the preparation ensemble are not to be averaged over crosslinking, because the network does not exist in the preparation state.

From the above example of nematic elastomers, it can be inferred that the structure of a polymer network in a randomly crosslinked system depends not only on the randomness inherent in the crosslinking process but also on the thermal fluctuations in the state of preparation.  Two macroscopically identical crosslinking processes using the identical protocol yield two distinct networks which are only {\it statistically\/} similar.  More importantly, two such crosslinking processes carried out under {\em different} preparation states yield {\em statistically different} network structures, and lead to different properties of the systems, even under the same conditions in the measurement state.  It therefore takes both the crosslinking methods and the preparation state to have a {\em full statistical description} of the structure of the random network in these materials.

In view of the three steps just mentioned, as well as their statistical nature, there are four types of physical quantities  that can be measured in a randomly crosslinked system:
\hfil\break\noindent
(1)~Thermodynamic averages in the
{\it preparation\/} state: 
$[A]_{\rm p}$,
$[A\,B]_{\rm p}$ etc.,
where the subscripts p stands for \lq\lq preparation\rlap.\rq\rq\thinspace\
\hfil\break\noindent
(2)~Thermodynamic averages in the
{\it measurement\/} state:
$\left[\,
\overline{
\langle A\rangle_{\rm m}
}\,
\right]_{\rm p}$,
$\left[\,
\overline{
\langle A B\rangle_{\rm m}
}\,
\right]_{\rm p}$ etc.
Note that these are to be averaged over both their crosslinking randomness {\it and\/} the thermal fluctuations in the preparation
state.
\hfil\break\noindent
(3)~Glassy correlations, which characterize sample-to-sample fluctuations in the measurement state:
$\left[
\overline{
\langle A\rangle_{\rm m}
\langle B\rangle_{\rm m}
}
\right]_{\rm p}$.
These quantities are important because the systems are random.
\hfil\break\noindent
(4)~Memory correlations: 
$\left[A\,
\overline{
\langle B\rangle_{\rm m}
}
\right]_{\rm p}$,
which characterize {\em cross-correlations between\/} the preparation and measurement states.  Memory correlations therefore characterize the dependence of system properties on their histories of formation.

The purpose of this Paper is to discuss various basic aspects of vulcanization theory, which provides a general methodology for the systematic calculation of all four of these types of diagnostic quantity.
The remainder of the Paper is organized as follows.  In Sec.~\ref{sec:model} we introduce a toy model of randomly linked anisotropic particles, as a starting point for vulcanization theory.
In Sec.~\ref{sec:Deam-Edwards} we discuss a generalized Deam-Edwards distribution of network structure, and average the free energy of the network over this distribution.  At this stage, we shall see a key theme, viz., that both the preparation and the measurement ensemble appear naturally, coupled to one another, in the effective replicated Hamiltonian.
In Sec.~\ref{sec:physical_quantities} we discuss in detail various physical quantities that can be calculated using the replicated partition function.   We also discuss the significance of the zeroth replica and the connection between replica limit and causality.  We also address the distinction between the original Deam-Edwards distribution and our generalization of it.
In Sec.~\ref{sec:order_parameter} we discuss the physical meaning of the vulcanization order parameter.
In Sec.~\ref{sec:Landau_theory} we study the mean-field approximation, and from it we derive the classical theory of rubber elasticity, as well as the neoclassical theory elasticity, appropriate for nematic elastomers.
%
%
Finally in Sec.~\ref{sec:conclusion}, we draw the concluding remarks by briefly reviewing three basic results about the heterogeneous nature of randomly crosslinked materials, and by indicating some possible future directions.





%
\section{Model of linked particles}
\label{sec:model}


To make our discussion concrete, we consider a liquid of particles, each of which may or may not be anisotropic.  The particles may, e.g., be entire polymers that we regard in a coarse-grained sense. Under appropriate conditions, the particles may participate in some kind of collective ordering, such as liquid crystalline alignment.  Each particle has certain number of functional sites at which the crosslinkers can act.  Before crosslinking, the spatial locations of all the functional sites of all the particles, which we index via $i$, are denoted by $\cv^0_i$.
We assume that the positions of these sites
$\cv^0 \equiv \{\cv^0_i, i = 1,\ldots,N\}$
completely specify the microscopic state of the liquid (i.e., there are no relevant degrees of freedom beyond these).  The liquid Hamiltonian (normalized by temperature) is denoted by $H^0_{\rm liq}(\cv^0)$.  The corresponding partition function for the liquid system (in the preparation state) is then given  (with the shorthand notation $D \cv^0 = \prod_{i=1}^N d \cv^0_i$) by
\be
Z^0_{\rm liq} = \int D \cv^0 \,
e^{- H^0_{\rm liq}(\cv^0) }.
\ee

\begin{figure}
	\centering
		\includegraphics[width=6cm]{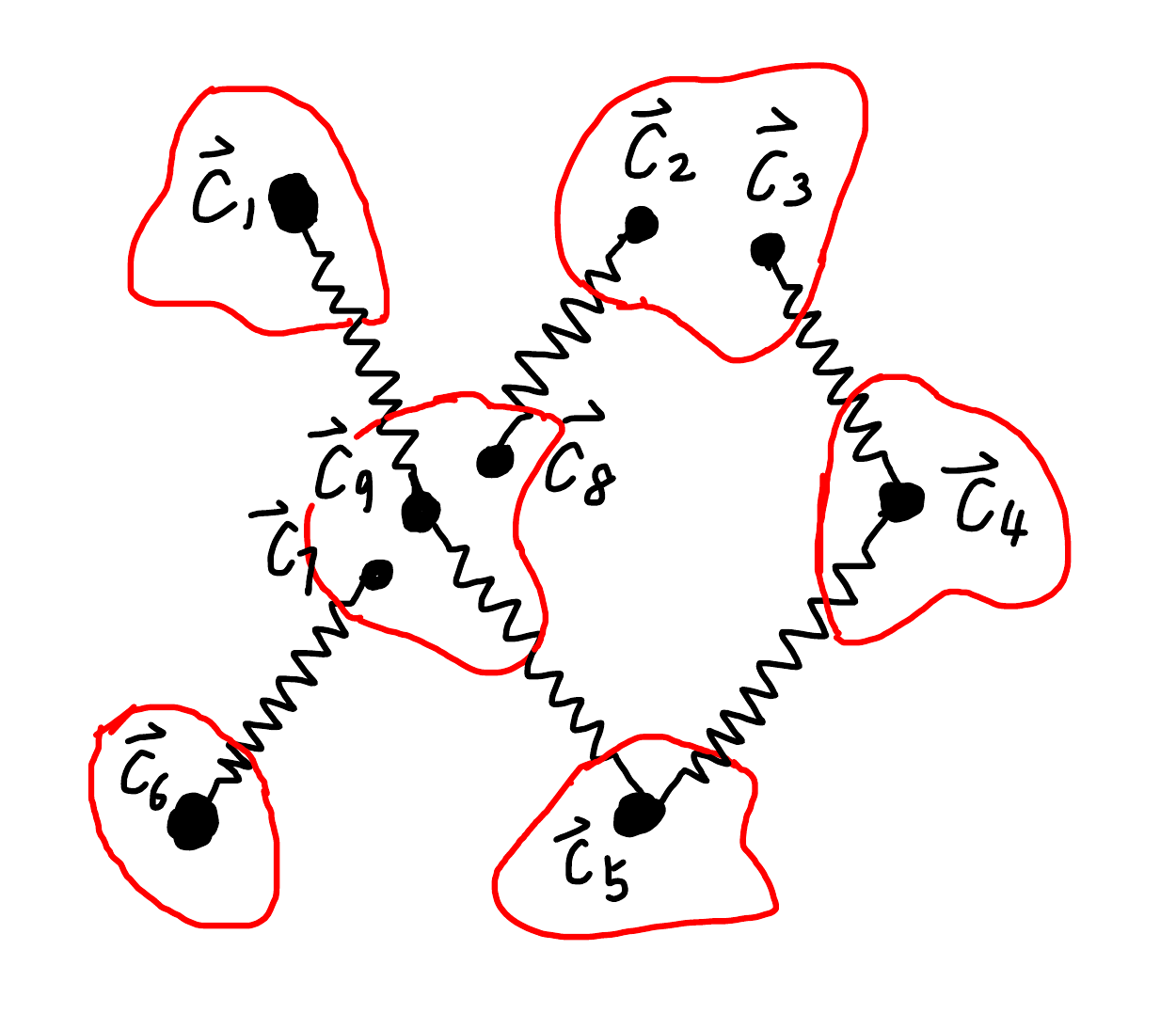}
	\caption{%
Collection of particles, schematically depicted as roughly spherical blobs, containing functional sites labeled $i$ and having spatial locations $\cv_i$.  Some number of crosslinkers (here, Gaussian springs) are introduced into the system; each of them can link a pair of functional sites.  The detailed interactions between the particles do not play a significant role in the present discussion.}
  \label{fig:cartoon-1}
\end{figure}

\begin{table}[!hbt]
\label{table:network-structure}
\caption{Two equivalent specifications of the network structure shown in Fig.~\ref{fig:cartoon-1}.}
\vspace{5mm}
\begin{tabular}{ll}
 \hline\hline  \\
 $\chi$\hspace{15mm}  &
 \begin{math} k_{(1,9)} = k_{(2,8)} = k_{(3,4)}
 = k_{(4,5)} = k_{(5,9)} = k_{(6,7)} = 1 \end{math},
 all other $k_{(i,j)} = 0$
  \vspace{5mm} \\
 $(M, \chi')$ &
 \begin{math}
M = 6, \quad \chi' = \{(i_e,j_e), e = 1,\ldots, 6 \} = \{(1,9), (2,8), (3,4), (4,5), (5,9), (6,7) \}
 \end{math}
  \vspace{3mm} \\ \hline
\end{tabular}
\end{table}

The crosslinkers are modeled as Gaussian springs, each connecting a pair of functional sites $(i, j)$ (for $i\leq j$ to avoid double counting).  The network structure is uniquely specified by the list of $N(N+1)/2$ \lq\lq link numbers\rq\rq\ $k_{(i,j)}$ that specify the number of springs linking the pair of sites $(i,j)$: $\chi=\{k_{(i,j)}\}$
\footnote{Note that we do permit a functional site to be connected to itself.  Even though for a lightly crosslinked system it would occur only rarely that a pair of functional sites would be multiply crosslinked (i.e., some $k>1$), we lose nothing by allowing such events and, as we shall see, it actually simplifies our theoretical analysis of the disorder average.  Also note  that we allow a spring to ``cross-link\rq\rq\ a functional site to itself, which corresponds to the case $i=j$.  This specification has some similarity to the grand canonical ensemble description of quantum gas statistics, in that it replaces a (difficult) constrained summation with an (easy) unconstrained summation.}.
Then, the summation over network structures  corresponds to summations over all (non-negative) integer values for $k_{(i,j)}$:
\be
\sum_{\chi} =
\sum_{k_{(1,1)} = 0}^{\infty}\sum_{k_{(1,2)} = 0}^{\infty}
\cdots
\sum_{k_{(N,N)} = 0}^{\infty}
\label{sum_k_ij}
\equiv \sum_{\{(i,j)\}}.
\ee

Another, equivalent, specification of the network structure is given as follows.
First, we specify the total number of springs $M$.  Second, we treat these $M$ springs as if they are distinguishable, and number them using an integer index $e = 1, \ldots, M$.  The $e^{\rm th}$ spring can be denoted by an integer pair $(i_e, j_e)$ (with $i_e \leq j_e$) which specifies the pair of functional sites it links together.  The network structure is then fully  {\it but non-uniquely\/} characterized by an integer $M$ and a list of $M$ integer pairs
$\chi' = \{(i_e, j_e) \}_{e=1}^M$.
Because all springs are identical, any arbitrary permutation of the $M$ pairs within the list $\chi'$ does not change the network structure.
We easily find the number of distinct $\chi'$ that correspond to the same network structure (which is uniquely determined by $\chi$), i.e., the ``degeneracy factor'':
\be
\frac{M!}{\prod_{(i,j)} k_{(i,j)}!}.
\label{degeneracy-factor}
\ee
The summation over all network structures can then also be represented as a summation over the crosslinker number $M$, combined with a sum over the list of $2M$ integers $\{i_e, j_e\}$,
\be
\sum_{\chi} =
\sum_{M=0}^\infty
\sum_{i_1, j_1} \cdots \sum_{i_M, j_M}.
\nonumber
\ee
Note, however, that the probability of a given network structure $P_{\chi'}$ must be divided by the degeneracy factor~(\ref{degeneracy-factor}) in order to correct for the over-counting.  We shall discuss this issue in detail below.
For an illustration of a particular network structure, see Fig.~\ref{fig:cartoon-1}.  For the two equivalent specification of this network structure, see Table \ref{table:network-structure}.

Let us assert that after crosslinking the system is brought to the measurement state, in which the microscopic locations of the sites are now labeled by $\cv_i$
(rather than  $cv^0_i$ of the preparation ensemble).
The interaction between pairs of sites linked by springs is modeled via quadratic terms in the separations between sites:
\be
\Delta H_{\chi} (\cv) =
\frac{1}{2 b^2}
\sum_{e=1}^M
\left\vert{\cv}_{i_e} - {\cv}_{j_e} \right\vert^2
= \frac{1}{2 b^2}
\sum_{(i,j)} k_{(i,j)}
\left\vert{\cv}_{i} - {\cv}_{j}\right\vert^2.
\label{deltaH-chi-def}
\ee
This potential energy increases without bound as any two linked sites are separated, and hence it keeps them close to each other at all times.
For later purposes, we also define the liquid-state partition function in the measurement state:
\be
Z_{\rm liq} = \int D \cv \, e^{- H_{\rm liq}(\cv) }.
\label{Z_liq-def}
\ee
The partition function and free energy of the {\em crosslinked} system are then given, respectively, by
\be
\label{eq:Zchi}
Z_{\chi} =
\int D{\bm{c}} \,
e^{-H_{\rm liq} (\cv) - \Delta H_{\chi} (\cv)  }, \quad
F_{\chi} =
-  T \ln Z_{\chi},
\ee
 both of which depend on the network structure $\chi$.  Throughout this paper, we shall set the Boltzmann constant to be unity.  Note that, generically, $H_{\rm liq} (\cv)$ (i.e., the liquid Hamiltonian in the measurement state) differs from its preparation-state counterpart $H^0_{\rm liq} (\cv)$.  We then average the free energy over realizations of the network structure $\chi$ (with probability $P_{\chi}$) to obtain
\be
\label{eq:disorderaverageF}
\overline{F} =
\sum\nolimits_\chi P_{\chi} F_\chi =
-  T \sum\nolimits_{\chi} P_{\chi} \ln Z_\chi \, .
\ee
Assuming that the selfaveraging property holds, this formula also gives the free energy of a typical sample.  The calculation (or modeling) of the probability $P_{\chi}$ for the network structure $\chi$ is the {\it sine qua non\/} for obtaining a proper statistical treatment of randomly crosslinked systems.   This we now discuss in detail.

\section{Generalized Deam-Edwards distribution of network structure}
\label{sec:Deam-Edwards}
\subsection{Statistics of connectivity between a pair of sites}
The crosslinking process that we shall consider is an instantaneous crosslinking process, as already mentioned previously. Many springs (crosslinkers) are introduced into the liquid system (in its preparation state), completely randomly and independently of one another.  We further assume that the extension of every spring $\mathbf{d}$ is distributed according to a Gaussian equilibrium Gibbs-Boltzmann factor with characteristic lengthscale $b$ [cf.~Eq.~(\ref{deltaH-chi-def})]:
\be P( \bm{d} ) \propto e^{ -| \bm{d} |^2/{2 b^2} }.
\label{GB-factor}\ee
In addition, we assume that each functional site $\cv^0_i$ has an ``effective region'' of radius $\epsilon$, centered around $\cv^0_i$.  If an end of a spring is introduced at some position $\xv$ that lies within the ``effective region'' around $\cv^0_i$, it is successfully linked to this functional site.  We assume that $\epsilon$ is smaller than half the minimal distance between any two functional sites, so that any end of a spring cannot be linked to two sites simultaneously.  This assumption does not limit the generality of our approach.


\begin{figure}
	\centering
		\includegraphics[width=7cm]{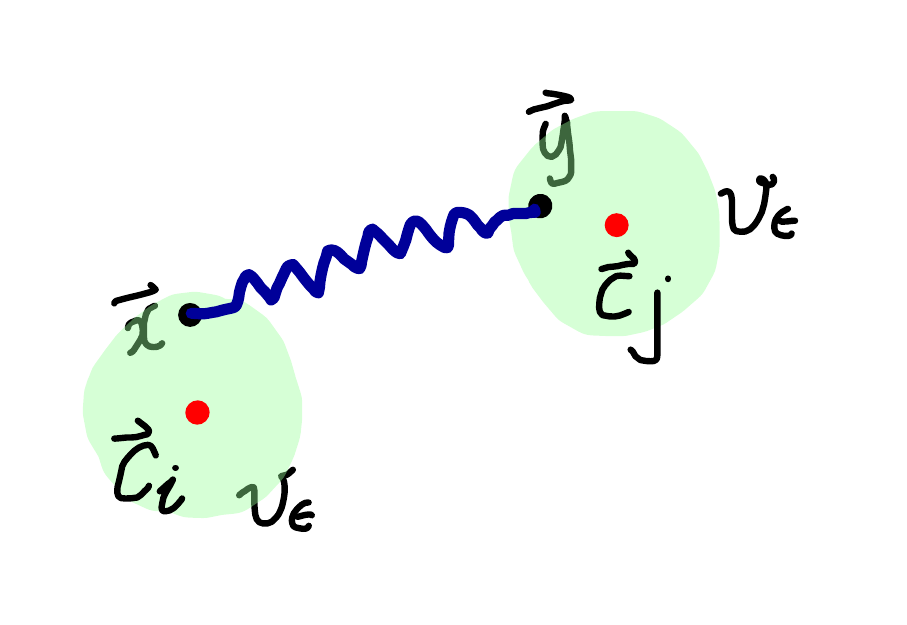}
	\caption{Two functional sites, $\cv_i$ and $\cv_j$, each with an ``effective region'' of radius $\epsilon$ and volume $v_{\epsilon}$.  Any end of a spring that is introduced into one such region is successfully linked to the site at the center of that region.}
  \label{fig:cartoon-2}
\end{figure}

Next, let $p_k(\cv^0_i,\cv^0_j)$ be the probability that a given pair of sites $\{\cv^0_i,\cv^0_j\}$ is linked by some integer number $k$ of springs.  If we assume that the crosslinkers are introduced into the system at random, independently and homogeneously, then $k$ would be Poisson-distributed:
\ba
p_{k}(\cv^0_i,\cv^0_j)
= C \, \frac{\hat{\mu}^k}{k!}\,
e^{- k\, \left|  \cv^0_i - \cv^0_j \right| ^2/2 b^2},
\label{p_k-1}
\ea
where we have introduced the dimensionless control parameter $\hat{\mu}$ for the crosslinking density~\footnote{$\hat{\mu}$ is proportional to $v_{\epsilon}^2$, where $v_{\epsilon}$ is the effective volume of each functional site, as defined above.}, and the constant $C$ is determined via overall normalization, see below.  
Note that $p_k(\cv^0_i,\cv^0_j)$ contains a factor proportional to the $k^{\rm th}$ power of Gibbs-Boltzmann factor~(\ref{GB-factor}), which characterizes the equilibrium distribution of each spring. It is convenient to define a (three-dimensional) soft delta function $\delta_b(\xv)$ in terms of a Gaussian spatial profile  having variance parameter $b^2$:
\ba
%
\delta_b(\xv) \equiv
( \sqrt{2\pi}/b )^{3}
\,e^{- \xv^2/2 b^2}.
\label{delta_b-def}
\ea
The function $\delta_b(\xv)$ tend to the Dirac delta function as $b\rightarrow 0$.  The linking probability~(\ref{p_k-1}) can then also be expressed in the following concise form:
\be
p_{k}(\cv^0_i,\cv^0_j) =
e^{- \tilde{\mu}  \, \delta_b(\cv^0_i - \cv^0_j)}
\frac{1}{k!}
\left(\,
\tilde{\mu} \, \delta_b(\cv^0_i - \cv^0_j) \,\right)^k,
\label{p_k-2}
\ee
where we have exchanged the crosslink density control parameter $\hat{\mu}$ for the parameter $\tilde{\mu}$, defined via
\be
\tilde{\mu} \equiv 
(b/ \sqrt{2\pi} )^{3}  \, {\hat{\mu}}.
\label{mu_tilde-def}
\ee
It is easy to check that Eq.~(\ref{p_k-2}) is properly normalized: $\sum_{k = 0}^{\infty} p_k(\cv^0_i,\cv^0_j)  =1$.


\subsection{Statistics of the connectivity of the network}
We assume that the liquid, prior to crosslinking, is in thermal equilibrium with respect to Hamiltonian $H^0_{\rm liq}(\cv^0) $, where $\cv^0 \equiv \{\cv^0_i\}$ denotes the locations of all $N$ functional sites in the preparation state~\footnote{This assumption is mainly made for simplicity, and is not essential for the development of vulcanization theory. Experimentally, it is also straightforward to crosslink a nonequiibrium state, such as the nematic-genesis polydomain nematic elastomers (N-PNE) prepared by Urayama {\it et al\/}.  To develop a vulcanization theory appropriate for this system, we would use the corresponding nonequilibrium probability density function.}. 
The joint probability density function for all $N$ sites (prior to crosslinking) to be at the locations $\{\cv^0_1,\ldots, \cv^0_N\}$ is then given by
\be
P(\cv^0) =
 \frac{1}{Z^0_{\rm liq} }\,
 e^{- H^0_{\rm liq}(\cv^0)},
\ee
where the normalizing partition function is given by
\be
Z^0_{\rm liq} = \int D \cv^0 \,
e^{- H^0_{\rm liq}(\cv^0)}.
\ee
Now recall that the network structure is characterized by a set of $N(N+1)/2$ stochastic variables \footnote{Note that a spring can connect a functional with itself. }, or link numbers, $\{k_{(i,j)}\}$, each of them characterizing the number of springs linking a particular pair $(i,j)$.
Given a liquid configuration $\cv^0$ prior to crosslinking, these link numbers are fully uncorrelated, and governed by the following conditional probability distribution:
\ba
&& P({\chi}|\cv^0) = \prod_{(i,j)}p_{k_{(i,j)}}(\cv^0_i, \cv^0_j)
\nonumber\\
&=& \prod_{(i,j)}
e^{- \tilde{\mu}  \, \delta_b(\cv^0_i - \cv^0_j)}
\frac{1}{k_{(i,j)}!}
\left( \tilde{\mu} \, \delta_b(\cv^0_i - \cv^0_j) \right)^{k_{(i,j)}},
\ea
where we have used the linking probability given by Eq.~(\ref{p_k-2}).  Making the further definition
\be
H_{\newsub} (\cv)
\equiv
- \tilde{\mu} \,
\sum_{i, j }
\delta_b({\cv}_i - {\cv}_j),
\ee
we can write this conditional probability as
\ba
P({\chi}|\cv^0) &=& e^{H_{\newsub}(\cv^0)}
\prod_{(i,j)} \frac{1}{k_{(i,j)}!}
\left( \tilde{\mu} \,
\delta_b(\cv^0_i - \cv^0_j) \right)^{k_{(i,j)}}
\nonumber\\
&= & e^{H_{\newsub}(\cv^0)}
\prod_{(i,j)} \frac{\hat{\mu}^{k_{(i,j)}}}{k_{(i,j)}!}
e^{- k_{(i,j)}\,|\cv^0_i - \cv^0_j|^2/2 b^2},
\nonumber\\
&=& e^{H_{\newsub}(\cv^0)-\Delta H_{\chi}(\cv^0)}
\prod_{(i,j)} \frac{\hat{\mu}^{k_{(i,j)}}}{k_{(i,j)}!} ,
\label{cond-prob-1}
\ea
where we have used Eq.~(\ref{delta_b-def}), Eq.~(\ref{mu_tilde-def}), and Eq.~(\ref{deltaH-chi-def}) in the above three equalities, respectively.
If we were to specify the network structure using the set $\chi' = \{(i_e,j_e), e = 1,\ldots, M\}$ instead of the set $\chi$, the corresponding conditional probability would be
\be
P(\chi'|\cv^0) =
e^{H_{\newsub}(\cv^0)}
\frac{1}{M!} \prod_{e = 1}^M \tilde{\mu} \, \delta_b(\cv^0_{i_e} - \cv^0_{j_e}),
\label{cond-prob-2}
\ee
where we have divided the probability by the proper degeneracy factor~(\ref{degeneracy-factor}), so as to cancel the over-counting.

The probability of obtaining the network structure $\chi$ can be arrived at by using the law of total probability as
\ba
&& P_{\chi} =
\sum_{\cv^0} P(\chi | \cv^0) P(\cv^0)
\nonumber\\
&=&
\frac{1}{Z^0_{\rm liq}}
\int D\cv^0 \,
e^{- H_{\rm liq}^0(\cv^0) + H_{\newsub}(\cv^0) -\Delta H_{\chi}(\cv^0)}\,
\prod_{(i,j)} \frac{\hat{\mu}^{k_{(i,j)}}}{k_{(i,j)}!}\,.
\label{P_chi}
\ea
Furthermore, by making the definition 
\ba
\tilde{Z}_{0,\chi}
&\equiv&
\int D \cv^0 \,
e^{-H_{\rm liq}^0(\cv^0) + H_{\newsub}(\cv^0)  -\Delta H_{\chi}(\cv^0)}\,,
\label{Z0chifirst}
\ea
we obtain
\ba
P_{\chi} =
\frac{\tilde{Z}_{0,\chi}}{Z^0_{\rm liq}}
\prod_{(i,j)} \frac{\hat{\mu}^{k_{(i,j)}}}{k_{(i,j)}!}.
\ea
Hence we see that, apart from a dimensionless factor, $P_{\chi}$ is the ratio of two partition functions.
Even though the variables $\cv^0_i$ (associated with preparation state) appear only as dummy variables, the integrations over them in Eq.~(\ref{Z0chifirst}) cannot be carried out explicitly.   Below, when we calculate the disorder average of free energy, we shall find that these variables $\cv^0_i$ {\it interact\/} with the corresponding  variables $\cv_i$ in the measurement ensemble.  It is this interaction that generates the ubiquitous memory effects in randomly crosslinked materials.

\subsection{Normalization of probabilities}

The conditional probability that there are precisely $M$ linking springs in the system can be obtained by summing Eq.~(\ref{cond-prob-2}) over all possible values taken by the indices $(i_e,j_e)$:
\ba
P(M|\cv^0) &=&
\sum_{i_1, j_1 = 1}^N  \sum_{i_2, j_1 = 2}^N
\cdots\sum_{i_M, j_M = 1}^N
P(\chi' | \cv^0)
\nonumber\\
&=& e^{H_{\newsub}(\cv^0)}
\frac{1}{M!}
\left(
- H_{\newsub}(\cv^0)
\right)^{M}.
\ea
It is elementary to check that the conditional probability $P(M|\cv_0)$ is properly normalized for any fixed $\cv_0$:
\be
\sum_{M = 0}^{\infty} P(M | \cv^0) = 1.
\ee
To check the normalization of the probability $P_{\chi}$, we sum Eq.~(\ref{P_chi}) 
over all possible network structures. By Eq.~(\ref{sum_k_ij}), this amounts to summing over
all possible linking numbers $\{ k_{(i,j)}$ \}.  Further using Eqs.~(\ref{deltaH-chi-def}) we arrive at 
\ba
\sum_{\chi} P_{\chi} &=&
 \frac{1}{Z^0_{\rm liq}} \int D\cv^0 \,
e^{- H_{\rm liq}^0(\cv^0) + H_{\newsub}(\cv^0)}
\sum_{\chi} e^{-\Delta H_{\chi}(\cv^0)} \,
\prod_{(i,j)} \frac{\hat{\mu}^{k_{(i,j)}}}{k_{(i,j)}!}
\nonumber\\
&= &\frac{1}{Z^0_{\rm liq}} \int D\cv^0 \,
e^{- H_{\rm liq}^0(\cv^0) + H_{\newsub}(\cv^0)}
\prod_{(i,j)} \sum_{k = 1}^{\infty}\frac{\hat{\mu}^k}{k!}
e^{ - k  \left| \cv_i - \cv_j \right|^2/2 b^2}. 
\ea
The summation and product can readily be calculated, leading to the factor
$\exp - H_X(\cv^0)$,
which precisely cancels the corresponding exponential factor.  Hence, as expected, we have the normalization condition
\be
\sum_{\chi} P_{\chi} = 1.
\ee

\subsection{Averaging over network structures}
We are now ready to compute the disorder-averaged free energy, Eq.~(\ref{eq:disorderaverageF}).  For this purpose, we use the following identity, which is commonly called the {\it replica trick\/} \footnote{Here, the overbar denotes an average over {\it network structure\/}, which is different from the average over crosslinker distributions (in real space) discussed previously.}:
\be
\overline {\ln X} = \lim_{n \rightarrow 0} \frac{1}{n} \ln \overline{X^n}.
\ee
We use this to rewrite the average free energy as
\ba
\overline{ F} &=&
-  T \lim_{n \rightarrow 0}
\frac{1}{n}
\ln \overline{Z_{\chi}^n}=
-  T \lim_{n \rightarrow 0}
\frac{1}{n}
\ln\sum_{\chi} P_{\chi} Z_{\chi}^n
\nonumber\\
&=&
-  T \lim_{n \rightarrow 0}
\frac{1}{n} \ln\sum_{\chi}
\left( \prod_{(i,j)} \frac{\hat{\mu}^{k_{(i,j)}}}{k_{(i,j)}!} \right)
\frac{1}{Z_{\rm liq}^0}\,\,
\tilde{Z}_{0,\chi}\,(Z_{\chi})^n.
\label{F-average-2}
\ea
We calculate the right hand side of Eq.~(\ref{F-average-2}) for positive integers $n$, and subsequently analytically continue $n$ to real values.  More specifically, we calculate $Z_{\chi}^n$ by replicating the variables $\{\cv_i\}$ a number $n$ times, to obtain an $n$-tuple
$(\cv^{1}_i, \cv^2_i,\ldots, \cv^n_i )$.
Together with the corresponding variable $\cv^0_i$ of the preparation ensemble, they form a $(1+n)$-tuple
$\hat{\cv}_i \equiv (\cv^{0}_i, \cv^1_i,\ldots, \cv^n_i )$.
The shorthand $\hat{\cv}$ has been extensively used for this $(1+n)$-tuple in the vulcanization theory literature.  Using Eqs.~(\ref{eq:Zchi}, \ref{Z0chifirst}), the product of partition functions
$Z_{0,\chi} ( Z_{\chi} )^n$ in Eq.~(\ref{F-average-2}) can be written in terms of functional integrals (with the shorthand notation $D\hat{\cv} \equiv \prod_{\alpha = 0}^n D \cv^{\alpha}$) as follows:
\ba
Z_{0,\chi}\,( Z_{\chi})^n =
\int D \hat{\cv} \,
\prod_{\alpha = 0}^n \,
e^{- \tilde{H}^0_{\rm liq} (\cv^0)
 - \sum_{\alpha = 1}^n H_{\rm liq} (\cv^{\alpha})
 - \sum_{\alpha = 0}^n \Delta H_{\chi}(\cv^{\alpha})
 }. 
 \label{Z_product_average}
\ea

Next, by using the definition Eq.~(\ref{deltaH-chi-def}) we easily see that 
\ba
 \sum_{\alpha = 0}^n \Delta H_{\chi}(\cv^{\alpha})
&=& \sum_{\alpha = 0}^n
\frac{1}{2 b^2}
\sum_{(i,j)} k_{(i,j)} \left| {\cv}_{i} - {\cv}_{j} \right| ^2
\nonumber\\
&=&\frac{1}{2 b^2}
\sum_{(i,j)} k_{(i,j)}
\left| {\hat{\cv}}_{i} - \hat{\cv}_{j} \right|^2
\nonumber\\
&\equiv& \Delta H_{\chi} (\hat{\cv}),
\ea
where we have used the notation
$\left| \hat{\cv}^{\alpha} \right|^2 \equiv \sum_{\alpha = 0}^n \left| \cv_i^{\alpha} \right|^2$.
We can now substitute Eq.~(\ref{Z_product_average}) back into Eq.~(\ref{F-average-2}),  exchange the order in which $\sum_{\chi}$ and $\int D \hat{\cv}$ are performed in Eq.~(\ref{F-average-2}), and sum over $\chi$ for a fixed configuration of $\hat{\cv}$.  The summation over the realizations of network structure can be calculated using the following result:
\ba
&&
\sum_{\chi} e^{- \Delta H_{\chi}(\hat{\cv})}
\left( \prod_{(i,j)} \frac{\hat{\mu}^{k_{(i,j)}}}{k_{(i,j)}!} \right)
\nonumber\\
&&\qquad=
 \sum_{\{k_{(i,j)} \}}
 \left( \prod_{(i,j)} \frac{\hat{\mu}^{k_{(i,j)}}}{k_{(i,j)}!} \right)
e^{- \frac{1}{2 b^2}
\sum_{(i,j)} k_{(i,j)} \left| {\hat{\cv}}_{i} - \hat{\cv}_{j} \right|^2}
\nonumber\\
&&\qquad=
 \prod_{(i,j)} \sum_{k}
\left( \frac{\hat{\mu}^{k}}{k!} \right)
e^{- \frac{1}{2 b^2}
k \left| {\hat{\cv}}_{i} - \hat{\cv}_{j} \right|^2}
\\
&&\qquad=
\prod_{(i,j)} \exp  \left(
\hat{\mu} \, e^{- \frac{1}{2 b^2}
k \left| {\hat{\cv}}_{i} - \hat{\cv}_{j} \right|^2}
\right)
\nonumber\\
&&\qquad=
\exp  \sum_{(i,j)}
{\mu} \, \delta_b(\hat{\cv}_i - \hat{\cv}_j)
\ea
where in the last equality, we have used the  definitions
\ba
{\mu} &\equiv& \hat{\mu} (\sqrt{2\pi}/b)^{3(1+n)}, \\
\delta(\hat{\xv}) &\equiv&
\prod_{\alpha = 0}^n \delta_b(\xv^{\alpha})
= \delta_b(\xv^0) \delta_b(\xv^1) \cdots \delta_b(\xv^n).
\label{delta_b-hat}
\ea
As we shall eventually take the replica limit, $n \rightarrow 0$, the difference between ${\mu}$ and $\tilde{\mu}$ [see Eq.~(\ref{mu_tilde-def})] can be ignored.

\vspace{2mm}
Combining all these results, we find that the disorder-averaged free energy is  given as follows:
\begin{subequations}
\ba
\overline{F} &=& -  T\, \lim_{n \rightarrow 0}  \frac{1}{n}
\, \ln\left( \frac{1}{Z^0_{\rm liq}} \, Z_{1+n}\right),
\label{F-average-3} \\
Z_{1+n} &=&  \int D\hat{\cv}
\, e^{-H_{1+n}(\hat{\cv}) },
 \label{Z_1+n}\\
H_{1+n}(\hat{\cv}) &=&
H_{\rm liq}^0(\cv^0) - H_{\newsub}(\cv^0)
+ \sum_{\alpha = 1}^n H_{\rm liq} (\cv^{\alpha})
 + H_{\newsub} (\hat{\bm{c}}), 
  \label{H_1+n}\\
 H_{\newsub} (\hat{\bm{c}}) &=&
- {\mu} \sum_{(i, j)}^N
\delta_b \left( \hat{\cv}_i - \hat{\cv}_j\right).
\label{H-X-def}
\ea
\end{subequations}
The Hamiltonian $H_{\newsub}[\hat{\bm{c}}] $ represents an effective short-range attraction between all pairs of functional sites in the replicated coordinate space $\hat{\bm{c}} = \{ \cv^{\alpha} \}_{n = 0}^n$, resulting from cross-linking.  This short-range interaction reduces to a delta function in the limit $b \rightarrow 0$.

\section{Physical quantities and correlation functions}
\label{sec:physical_quantities}
We shall use the notation $\langle \Psi(\hat{\cv}) \rangle_{1+n}$ to indicate averages of physical quantities with respect to the weight in the replicated partition function Eq.~(\ref{Z_1+n}):
\be
\langle  \Psi(\hat{\cv})  \rangle_{1+n} =
\frac{\int D \hat{\cv} \,
\Psi(\cv^0,\cv^1,\cdots, \cv^n)\, e^{- H_{1+n}} }
{\int D \hat{\cv}  \, e^{- H_{1+n}} }.
\label{ave_1+n-def}
\ee
When there is no risk of confusion, we shall neglect the subscript $1+n$ on the average and simply write $\langle \Psi(\hat{\cv}) \rangle$.  It is understood that at the end of the calculation of such quantities the replica limit $n \rightarrow 0$ should always be taken.

In the replicated Hamiltonian $H_{1+n}$, Eq.~(\ref{H_1+n}), only the last term, $H_{\newsub}(\hat{\cv})$,  couples together various replicas.  It can be understood as a short-range attraction between all pairs of the $1+n$ {\em replicated} positions $\hat{\cv} = (\cv^0,\cv^1,\ldots,\cv^n)$.
The strength of this attraction is controlled by the density of crosslinking ${\mu}$.  If ${\mu} = 0$, i.e., the system is not crosslinked, $H_{\newsub} = 0$, and $H_{1+n}$ reduces to a sum of liquid Hamiltonians, one for each replica:
\ba
H_{1+n} &\rightarrow&
{H}^0_{\rm liq} (\cv^0)
+ \sum_{\alpha = 1}^n H_{\rm liq} (\cv^{\alpha}),\nonumber\\
Z_{1+n} &\rightarrow& Z^0_{\rm liq} \cdot \left(Z_{\rm liq}\right)^n, \nonumber\\
\overline{F} &\rightarrow& -  T \ln Z_{\rm liq}.  \nonumber
\ea
Variables in distinct replicas then become completely decoupled from one another, and the replicated partition function $Z_{1+n}$ describes the equilibrium liquid in the preparation state together with $n$ copies of equilibrium liquid in the measurement state, with no interactions amongst these liquids.


For systems with crosslinks, ${\mu} \neq 0$ and so there are nonvanishing correlations between the variables $\cv^0$ and the other $\cv^{\alpha}$.  These correlations characterize the dependence of the properties of these systems on the history of formation. Let us look systematically at the various physical quantities that can be constructed within the framework of an ($1+n$)-replica theory, and relate them to the suite of correlators involving the preparation and measurement ensembles discussed in the Introduction.  Below, we use the shorthand notation
$A^{\alpha} \equiv A(\cv^{\alpha})$
for physical quantities, and we frequently omit the subscript $1+n$ on various expectation values, all to ease the notation.
\begin{enumerate}
\item One-replica quantities:
    $\langle A^0 \rangle$ and $\langle A^{\alpha} \rangle$  ($\alpha \neq 0$) are average physical quantities in the preparation state and in the measurement state, respectively.   In the notation used in Sec.~\ref{sec:intro}  they are written as $\left[ A \right]_{\rm p}$ and $\left[ \overline{\langle A \rangle_{\rm m}} \right]_{\rm p}$, respectively.

There are also correlation functions in a single replica:
$\langle A^0\,B^0 \rangle = [ A B ]_{\rm p}$
and
$\langle A^{\alpha}\,B^{\alpha} \rangle =
\left[ \overline{\langle A\,B \rangle_{\rm m}} \right]_{\rm p}$ 
                 ($\alpha \neq 0 $), which quantify correlations in the preparation state and in the measurement state, respectively.

\item Two-replica quantities:
    $\langle A^{\alpha}\,B^{\beta}  \rangle =
    \overline{ \langle A \rangle_{\rm m} \langle B \rangle_{\rm m}} $
    (for $\alpha \neq \beta; \alpha, \beta \neq 0$)
    are nonvanishing in the presence of the quenched disorder arising from random crosslinking, and
    thus give the glassy correlations in the measurement state.   The order parameter for the random solid state is an especially important example of such a multi-replica quantity.

Memory correlation functions involving $0$-th replica only once,
$\langle A^0  B^{\alpha} \rangle_c =
\left[ A  \, \overline{ \langle B \rangle_{\rm m} } \right]_{\rm p}$
(for $ \alpha \neq 0$)
measure the cross correlation between the preparation state and the measurement state.   These memory effects are a hallmark property of randomly crosslinked materials.   Vulcanization theory provide the unique framework via which they can be systematically calculated.



\end{enumerate}

In the setting of liquid crystalline elastomers, the memory correlators are the quantities that can best distinguish systems prepared under distinct physical conditions, e.g., isotropic genesis nematic elastomers (IGNEs) and nematic genesis nematic elastomers (NGNEs).  Various correlation functions, expressed in both replica notation and physical notation, are summarized in Table~\ref{table:correlators}.

\begin{table}[!hbt]
\label{table:correlators}
\caption{Various correlation functions in vulcanization theory, showing
the correspondence between physical notation and replica notation.
(Note that in the table we have $0 \neq \alpha \neq \beta  \neq 0$.)\thinspace\
In the replica notation, all averages are defined by Eq.~(\ref{ave_1+n-def}). }
\vspace{5mm}
\begin{tabular}{ccccc}
 \hline\hline  \\ \vspace{3mm}
 & \pbox{3cm}{Preparation \\ correlation}  \hspace{5mm}
 & \pbox{3cm}{Measurement \\ correlation} \hspace{5mm}
& \pbox{3cm}{ Glassy \\ correlation} \hspace{5mm}
 & \pbox{3cm}{Memory \\ correlation}
\\  \hline \\ \vspace{5mm}
\pbox{3cm}{Physical \\ notation}
& $\left[ A B \right]_{\rm p} $
& $\left[  \overline {\langle A B \rangle_{\rm m}} \right]_{\rm p}$
& $\left[ \overline {\langle A \rangle_{\rm m} \langle B \rangle_{\rm m}} \right]_{\rm p}$
& $\left[ A  \overline{ \langle B \rangle_{\rm m}} \right]_{\rm p}$
\\ \hline \\ \vspace{3mm}
\pbox{3cm}{Replica \\ notation}
& $\langle A^0 B^0 \rangle$
& $\langle A^{\alpha} B^{\alpha} \rangle$
& $\langle A^{\alpha} B^{\beta} \rangle$
& $\langle A^0 B^{\alpha}\rangle$
\\ \hline
\end{tabular}
\end{table}

\subsection{Significance of the zeroth replica}
We can use $Z_{1+n}$ with appropriate source terms to calculate the average of a physical quantity $A(\cv^0)$ ($\equiv A^0$) that only depends on the degrees of freedom $\cv^0$ of the preparation ensemble.
As replicas $1,\ldots, n$ play no role in such a calculation, we can take the replica limit $n \rightarrow 0$ {\it before\/} the calculation of the expectation value.  This is equivalent to deleting all variables $\cv^{\alpha}$ for $\alpha = 1, \ldots, n$ from the replicated Hamiltonian~(\ref{H_1+n}).  In particular, we have
\ba
H_{\newsub}(\hat{\cv}) &\rightarrow& H_{\newsub}(\cv^0), \nonumber\\
H_{1+n}(\hat{\cv}) &\rightarrow& H_{\rm liq}^0(\cv^0). \nonumber
\ea
We also have to delete the integrals over $\cv^{\alpha}$ (for $\alpha = 1,\ldots,n$) in the functional integrals~(\ref{Z_1+n}) and (\ref{ave_1+n-def}).  Therefore, we have
\ba
\langle A^0 \rangle  = \frac{\int D \hat{\cv} \,
A^0\, e^{- H_{1+n}} }
{\int D \hat{\cv}  \, e^{- H_{1+n}} } =
\frac{\int D \cv^0 \,
A^0 \, e^{- H^0_{\rm liq} (\cv^0) } }
{\int D \cv^0  \, e^{- H^0_{\rm liq} (\cv^0) } }
= \langle A \rangle_{\rm liq}^0
\equiv [ A ]_{\rm p}.
\label{ave-A_P}
\ea
As a result, $\langle A^0 \rangle$ is indeed the average in the preparation state, as expected.


\subsection{Replica limit and causality}
\begin{figure}
	\centering
		 \includegraphics[width=6cm]{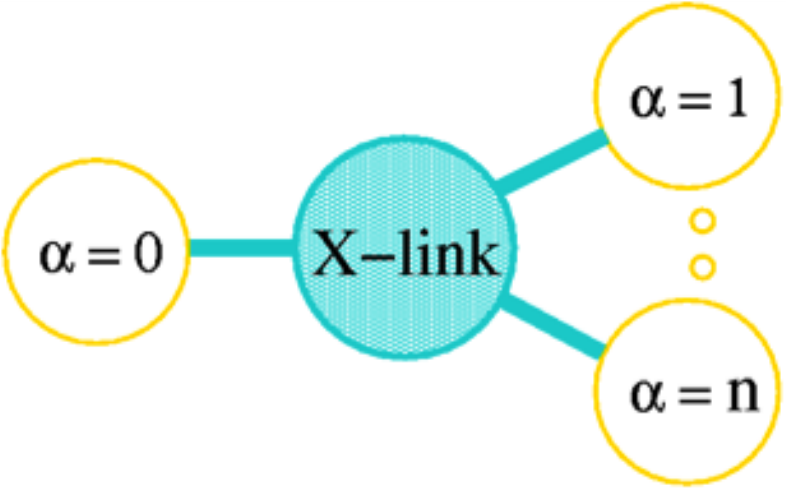}
	\caption{Cartoon showing the logical relations between different replicas: the preparation ensemble and $n$ copies of the measurement ensemble are coupled to one another via the crosslinking Hamiltonian $H_{\newsub}(\hat{\cv})$.   The relative weight of measurement ensemble vanishes in the replica limit, thus protecting the causality principle. }
  \label{fig:crosslink-replicas}
\end{figure}

In the replicated partition function, Eq.~(\ref{Z_1+n}), there is one copy of the preparation ensemble (the $0$-th replica), and $n$ copies of the measurement ensemble.  In Fig.~\ref{fig:crosslink-replicas} we show  a cartoon that shows schematically the relationship between various replicas.  All $1+n$ replicas interact with one another through the last term in Eq.~(\ref{H_1+n}), which is linear in the crosslinking density.  Note that there are $n$ copies of the measurement ensemble (i.e., replicas $1,\ldots, n$) but only one copy of the preparation ensemble (i.e., replica 0).  In general, owing to the interactions, the preparation and measurement ensembles can mutually influence each other.  For example, if we tune some parameter $J^0$ in fluid Hamiltonian $H^0_{\rm liq}$ of the preparation state [see Eq.~(\ref{H_1+n})], there will be changes in the physical quantities in the measurement state:
\be
\frac{\partial}{\partial J^0}
\left[\overline{\langle A
\rangle_{\rm m}}\right]_{\rm p}
\neq 0.
\ee
This is, of course, entirely reasonable, as the properties of polymer networks {\it do\/} depend on their history of formation.  Reciprocally, however, if we change a parameter $J$ in the {\em measurement state}, might there, at least in principle, also be responses of physical quantities in the preparation state, e.g. 
\be
\frac{\partial}{\partial J}
[A ]_{\rm p}
\neq 0 \quad (\bm{?}).
\ee
Such a response would be entirely {\em unphysical}, as it would violate the principle of causality.  In the framework of vulcanization theory, the causality principle is rescued via the replica trick: as $n \rightarrow 0$, the  weight of the measurement ensemble relative to the preparation ensemble tends to zero, and therefore the former can have no quantitative influence on the latter.

An explicit proof of this result is simple.  Let the parameter $J$ couple linearly to some quantity $\Phi(\cv)$ in the $(1+n)$-replica Hamiltonian $H_{1+n}$; this leads to the change
\be
H_{1+n} \rightarrow H_{1+n}
- J \sum_{\alpha = 1}^n \Phi(\cv^{\alpha}).
\nonumber
\ee
Substituting this back into the expectation value Eq.~(\ref{ave-A_P}), taking the derivative with respect to $J$, and finally setting $J = 0$, we obtain
\be
\frac{\partial [ A ]_{\rm p} }{\partial J}
= \sum_{\alpha = 1}^n
\langle A^0 \Phi(\cv^{\alpha}) \rangle
= n \langle A^0 \Phi(\cv^{1}) \rangle,
\label{A_p-J}
\ee
where in the last equality we have assumed that permutation symmetry amongst replicas 1 to $n$ remains intact.  The right hand side of Eq.~(\ref{A_p-J}) therefore vanishes in the replica limit, i.e., $n \rightarrow 0$.  Even in the presence of the spontaneous breaking of this replica permutation symmetry, the right hand side of Eq.~(\ref{A_p-J}) would even be proportional to $n$, and hence would vanish in the replica limit.




\subsection{Deam-Edwards revisited}
In the replicated Hamiltonian~(\ref{H_1+n}), the parameters of the preparation state only appear in $H^0_{\rm liq}$, whereas the parameters of the measurement state only appear in $H_{\rm liq}$.  These two types of parameters can be separately controlled in experiments, but they jointly determine the statistical physics of randomly crosslinked materials.   Let us consider a rather special case in which $H^0_{\rm liq} = H_{\rm liq}$, i.e., a system that is prepared and measured under identical conditions.  What appears inside the $(1+n)$-replica Hamiltonian~(\ref{H_1+n}) is, however, $H_{\rm liq}^0(\cv^0) - H_{\newsub}(\cv^0)$ instead of just ${H}^0_{\rm liq}(\cv^0)$:
\be
H_{1+n}(\hat{\cv}) =
H_{\rm liq}(\cv^0) - H_{\newsub}(\cv^0)
+ \sum_{\alpha = 1}^n H_{\rm liq} (\cv^{\alpha})
 + H_{\newsub} (\hat{\bm{c}}).
  \label{H_1+n-ours}
\ee
Consequently, the replica Hamiltonian $H_{1+n}$ is {\em not} invariant under permutations of the replicas that mix the $0$-th and the other replicas.  Hence, the average of a quantity $A$ in the preparation state is generically different from its average in the measurement state:
\be
[ A ]_{\rm p} = \langle A^0 \rangle
\neq \langle A^{\alpha} \rangle
= \overline{ \langle A \rangle_{\rm m}}.
\nonumber
\ee
This is completely reasonable, as the random network structure {\it does\/} affect physical quantities in the measurement state but cannot not influence the preparation state.  One simple example involves the nematic correlations in liquid crystalline elastomers that have been crosslinked in isotropic state.  The random polymer network tends to disorder the nematic degrees of freedom, and therefore reduces the nematic correlations.  

Historically, Deam and Edwards~\cite{Deam-Edwards-1976} chose a particular probability distribution $P_{\chi}$ for the network structure given by
 \ba
P_{\chi}^{\rm DE} =
\frac{\tilde{Z}_{0,\chi}^{\rm DE}}{Z^0_{\rm liq}}
\prod_{(i,j)} \frac{\hat{\mu}^{k_{(i,j)}}}{k_{(i,j)}!},
\ea
where
\ba
\tilde{Z}_{0,\chi}^{\rm DE} = \int D \cv^0 \,
e^{-H_{\rm liq}^0(\cv^0)  -\Delta H_{\chi}(\cv^0)}.
\label{Z0chi}
\ea
The corresponding $(1+n)$-replica Hamiltonian is then given by
\ba
H_{1+n}^{\rm DE}(\hat{\cv}) &=&
H_{\rm liq}(\cv^0)
+ \sum_{\alpha = 1}^n H_{\rm liq} (\cv^{\alpha})
 + H_{\newsub} (\hat{\bm{c}}) , \nonumber\\
 &= &  \sum_{\alpha = 0}^n H_{\rm liq} (\cv^{\alpha})
 + H_{\newsub} (\hat{\bm{c}}) .
  \label{H_1+n-DE}
\ea
in which all $1+n$ replicas appear symmetrically.  The resulting $1+n$ replica theory then has an enlarged $S_{1+n}$ permutation symmetry, rather than the $S_n$ permutation symmetry that is mandatory for replica theories. The Deam-Edwards choice of disorder distribution is appealing, in that it simplifies the analysis considerably.  However, it also leads to the unnatural consequence that the averages of arbitrary physical quantities in the preparation state are equal to their  counterparts in the measurement state.  The $S_{1+n}$ permutation in the Deam-Edwards theory is therefore not mandatory, as first pointed out by Broderix et al.~\cite{Zippelius-EPJB-2002}.   The explicit breaking of the $S_{1+n}$ permutation symmetry down to $S_n$ permutation symmetry is physically natural, and opens the way to a systematic analysis of the interrelationship between randomly crosslinked materials and their histories of formation.  This is a valuable and important element of the physics contained in vulcanization theory.


\subsection{Comparison with spin glasses and other disordered systems}
We end this section with a brief discussion of the difference between the vulcanization model and certain well-known spin glass models.  Consider, e.g., the Edwards-Anderson model~\cite{Edwards-Anderson-SpinGlass-1975}:
\be
H = - \sum_{\langle i,j\rangle}
J_{ij} S_i S_j. \nonumber
\ee
The quenched random variables $\{J_{ij}\}$ are typically assumed to have independent Gaussian statistics with vanishing means, and therefore averaging over them (after application of the replica method) is elementary.  The final, replicated theory reads:
\be
H_R = - \frac{1}{2} \sigma_J^2 \sum_{\langle i,j\rangle}
\sum_{\alpha, \beta} S^{\alpha}_i S^{\alpha}_j
S^{\beta}_i S^{\beta}_j,
\nonumber
\ee
and contains $n$ replicas of the annealed degrees of freedom (i.e., the spins). The variance of the quenched variables $\sigma_J$ appears only as a parameter in the replicated Hamiltonian.   {\it The quenched variables $\{J_{ij}\}$ have been integrated out, and therefore do not show up in the replica theory.}\thinspace\ By contrast, in  vulcanization it is not just expedient but indeed manifestly physical to employ an equilibrium preparation state to {\em generate} the statistics for the network structure.  (For vulcanization, the quenched random variables are the network structure.)\thinspace\  Furthermore, correlations between the preparation state variables and the measurement state variables encode the permanent memory effects, and these are experimentally measurable.  Finally, we note that the Edwards-Anderson model represents a considerable idealization of disordered systems in real world.  In general, the physics of most disordered systems does depend on the history of formation.  Hence, the proper statistical modeling of such systems requires a separate statistical ensemble.

\section{Order parameter for vulcanization}
\label{sec:order_parameter}
The crosslinking-induced interaction~(\ref{H-X-def}) couples distinct functional sites.  It can be decoupled using the standard Hubbard-Stratonovich transformation.  To proceed, we first notice that Gaussians are closed under convolution.  Therefore, the following identity about the $(1+n)d$-dimensional soft delta function $\delta_b(\xh)$ [defined in Eqs.~(\ref{delta_b-def},\ref{delta_b-hat})] can readily be established:
\be
\delta_b(\xh -\yh) =
\int d \zh \, \delta_{b'} (\xh - \zh) \,
\delta_{b'} (\zh - \yh),
\quad\quad b' = b/\sqrt{2}.
\nonumber
\ee
Equation~(\ref{H-X-def}) can then be written as
\ba
 H_{\newsub} (\hat{\bm{c}}) &=&
- {\mu} \int d \hat{\bm{z}}
 \sum_{i=1}^N \delta_{b'}
  \left( \hat{\cv}_i - \hat{\bm{z}}\right)
\sum_{j=1}^N \delta_{b'}
  \left( \hat{\cv}_j - \hat{\bm{z}}\right)
  \nonumber\\
&=& - {\mu} \int d \hat{\bm{z}} \,
\omega(\zh,\hat{\bm{c}})^2,
\nonumber
\ea
where
\be
\omega(\zh,\hat{\bm{c}}) \equiv
 \sum_{i=1}^N \delta_{b'}
  \left( \hat{\cv}_i - \hat{\bm{z}}\right)
\label{collective-field-def}
\ee
is the microscopic collective density of functional sites, coarse grained by the Gaussian wave packet $\delta_{b'}(\xh)$.  It depends explicitly on the microscopic configuration of the replicated system.  Our definition of the collective field differs from those in the other literature on vulcanization theory \cite{Vulcan_Goldbart} via a multiplicative factor of $N$.   We can now introduce an auxiliary field $\Omega(\xh) \equiv \Omega(\xv^0,\ldots, \xv^n)$ via the following Hubbard-Stratonovich transformation:
\ba
&& e^{-H_{\newsub} (\hat{\bm{c}})}
= e^{\, {\mu} \int d \hat{\bm{z}} \,
\omega(\zh,\hat{\bm{c}})^2}
\nonumber\\
&=&
C  \int D \Omega(\xh) \,
e^{- {\mu} \int d\xh \left( \Omega(\xh)^2
- 2 \, \omega(\xh) \Omega(\xh)\right)},
\nonumber
\ea
where $C$ is a trivial constant that will be neglected henceforth.  Insertion of this result into Eq.~(\ref{Z_1+n}) leads to
\ba
Z_{1+n} &=&
\int
D \hat{\cv} \int D \Omega(\xh) \,
e^{ - H^0_{\rm lid}[\cv^0] + H_X[\cv^0]
- \sum_{\alpha = 1}^n H_{\rm liq}^{\alpha}[\cv^{\alpha}]
- {\mu} \int d\xh \, \Omega(\xh)^2
+ 2 \,{\mu}  \int d\xh \,
 \omega(\xh)\, \Omega(\xh)
}
\nonumber\\
&=&
 \int D \Omega(\xh) \,
 e^{
 - {\mu}
\int d\xh \, \Omega(\xh)^2
}
\int D \hat{\cv} \,
e^{ - H^0_{\rm lid}[\cv^0] + H_X[\cv^0]
- \sum_{\alpha = 1}^n H_{\rm liq}^{\alpha}[\cv^{\alpha}]
+ 2\,{\mu}  \int d\xh \,
 \omega(\xh)\, \Omega(\xh)
}.
\label{Z_1+n-Omega}
\ea
Let us further define a modified liquid state partition function via
\be
\tilde{Z}_{\rm liq}^0 \equiv \int D\cv^0 \,
e^{- H^0_{\rm liq}[\cv^0] + H_X[\cv^0]},
\ee
and the average $\langle \, \cdots \, \rangle^0_{1+n}$ as
\be
\langle \Psi(\hat{\cv}) \rangle^0_{1+n} \equiv
\frac{1}{\tilde{Z}_{\rm liq}^0 \cdot \left( Z_{\rm liq}\right)^n}
\int D \hat{\cv} \, \Psi(\hat{\cv})\,
e^{ - H^0_{\rm lid}[\cv^0] + H_X[\cv^0]
- \sum_{\alpha = 1}^n H_{\rm liq}^{\alpha}[\cv^{\alpha}]
}.
\ee
Then we can rewrite Eq.~(\ref{Z_1+n-Omega}) as
\be
Z_{1+n} =
\tilde{Z}_{\rm liq}^0 \cdot \left( Z_{\rm liq}\right)^n \cdot
\int D \Omega(\xh) \, e^{- H_{\rm eff}(\Omega)},
\label{Z_1+n-Omega-new}
\ee
where the effective Hamiltonian $H_{\rm eff}(\Omega)$ is defined as
\ba
H_{\rm eff}(\Omega) = 
{\mu} \int d\xh \, \Omega(\xh)^2
- \ln \left\langle
e^{ 2\,{\mu}  \int d\xh \,
 \omega(\xh) \Omega(\xh) }
\right\rangle_{1+n}^0.
\label{H_eff-def}
\ea

We can now define the average of functions of $\Omega$ using the equilibrium partition function associated with $H_{\rm eff}(\Omega)$:
\be
\langle \Upsilon(\Omega)\rangle_{\rm eff} \equiv
\frac{\int D \Omega \, \Upsilon(\Omega) \,
e^{- H_{\rm eff}(\Omega)}
}{
\int D \Omega \,
e^{- H_{\rm eff}(\Omega)}}.
\label{ave-field-theory}
\ee
The following identity is a well-known property of the Hubbard-Stratonovich transformation:
\ba
 \left \langle  \Omega(\xh) \right \rangle_{\rm eff}
=\left \langle \omega(\xh,\hat{\bm{c}}) \right \rangle_{1+n}
= \left\langle \sum_{i=1}^N
\delta_{b^{\prime}}(\xv^0 - \cv_i^0)
\cdots
\delta_{b^{\prime}}(\xv^n - \cv_i^n)
\right\rangle_{1+n} ,
\label{Omega-def}
\ea
which shows the physical significance of the collective field $\Omega(\xh)$, namely that its average is the joint probability density function of the following $1+n$ events: in the preparation state there is a particle at some point $\xv^0$, whereas in $n$ independent measurements on the measurement state {\it the same particle\/} is found at positions $\xv^1, \ldots,\xv^n$, respectively.  Using the physical notation discussed in the Introduction, this probability density can be expressed as
\be
 \left \langle  \Omega(\xh)
 \right \rangle_{\rm eff}
 =
 \sum_{i=1}^{N} \left[
\delta_{b^{\prime}}(\xv^0- \cv_i)
 \overline{
  \langle
\delta_{b^{\prime}}(\xv^1- \cv_i)
\rangle_{\rm m}
\cdots
\langle
\delta_{b^{\prime}}(\xv^n- \cv_i)
\rangle_{\rm m} }
\right]_{\rm p},
\label{Omega-def-physical}
\ee
where, as usual, the subscripts p and m respectively denote the preparation and measurement states, and the overline indicates an average over realizations of  crosslinkers.  $\Omega(\xh)$ encodes  information about the localization of particles in the gel phase, and consequently is called {\em  the vulcanization order parameter}.  The definition of $\Omega(\xh)$ in this work differs from that in some of the other vulcanization literature via a trivial additive constant and a multiplicative constant $N$.

\subsection{One replica sector and marginal distributions}
We can integrate out the $n$ variables $\xv^1,\ldots, \xv^n$ in the order parameter~(\ref{Omega-def}),  leaving only one variable $\xv^0$, and thereby obtain a one-replica distribution
\be
\langle \Omega^{0}(\xv^{0}) \rangle_{\rm eff}
\equiv
 \int \prod_{\alpha = 1}^n d \xv^{\alpha}
\langle \Omega(\hat{\xv}) \rangle
= \left \langle \sum_{i=1}^{N}
\delta_{b^{\prime}}(\xv^0 - \cv^0_i)
\right \rangle_{1+n}
 = \left[ \rho(\xv^0)  \right]_{\rm p} .
 \label{Omega^0}
\ee
The resulting object is the average number-density of functional sites in the {\em preparation state}, which is manifestly an intensive quantity.  At the coarse-grained level, it can also be understood as the {\em average particle density} of the liquid prior to crosslinking.  The fluctuation
$\delta \Omega^{0}(\xv^{0}) = \Omega^{0}(\xv^{0}) - \langle \Omega^{0}(\xv^{0}) \rangle_{\rm eff}$
is also linearly related to the fluctuation of particle density in the preparation state.  Similarly, we can obtain the $\alpha$-th replica distribution for $\alpha \neq 0$:
\be
\langle \Omega^{\alpha}(\xv^{\alpha}) \rangle_{\rm eff}
\equiv
\int \prod_{\beta \neq \alpha}^n d \xv^{\beta}
\langle \Omega(\hat{\xv}) \rangle
= \left \langle \sum_{i=1}^{N}
\delta_{b^{\prime}}(\xv^{\alpha} - \cv^0_i) \right \rangle_{1+n}
= \left[ \overline{\langle \rho(\xv^{\alpha}) \rangle_{\rm m} }\right]_{\rm p} ,
\label{Omega^alpha}
\ee
which describes average particle density in the measurement state.  This relationship between the vulcanization order parameter and the particle densities in the preparation and measurement states 
is the primary reason that we choose the particular normalization of the order parameter field in Eq.~(\ref{collective-field-def}).  

For swollen gels and elastomers there can be large density fluctuations in the system (either quenched or thermal).   These fluctuations, characterized by correlations of the one-replica distributions $\Omega^{\alpha}(\xv^{\alpha})$, may strongly influence the elasticity of system.  For unswollen gels and elastomers, the mean particle density at long enough lengthscales is usually uniform and its fluctuations are negligibly small.  The one-replica part of free energy then contains no interesting physics, besides the stabilizing of the density fluctuations.

Finally, by integrating out $n-1$ variables, we obtain the two-replica correlator
\be
\langle \Omega^{\alpha \beta}(\xv^{\alpha} - \xv^{\beta}) \rangle_{\rm eff}
= \int \prod_{\gamma \neq \alpha,\beta}^n d \xv^{\gamma}
\langle \Omega(\hat{\xv}) \rangle,
\label{Omega-bar-0}
\ee
which encodes the system average of the correlation of the position fluctuations of a site in two independent measurements.  Macroscopic translational symmetry dictates that this function depend only on the difference between two coordinates, $\xv^{\alpha}$ and $\xv^{\beta}$.  Generically, it contains less information than the full order parameter $\Omega(\xh)$.


\section{Landau theory and saddle-point approximation}
\label{sec:Landau_theory}
Eqs.~(\ref{Z_1+n-Omega-new}, \ref{ave-field-theory}) provide a formal approach to calculate all relevant physical quantities.  Our problem is therefore reduced to the calculation of a functional integral over $\Omega(\xh)$.  At the saddle-point level we approximate the disorder-averaged free energy~(\ref{F-average-3}) using the minimum of the effective Hamiltonian $H_{\rm eff}[\Omega]$, Eq.~(\ref{H_eff-def}).  Even after making this approximation, the problem remains difficult, because we do not know the effective Hamiltonian~(\ref{H_eff-def}) exactly, let alone its minimum.  We therefore proceed to expand the exponential in Eq.~(\ref{H_eff-def}) in a Taylor series in $\Omega$.   At this stage, it is convenient to shift the order parameter~(\ref{Omega-def}) by a constant $N/V^{1+n}$ (where $N$ is the total number of functional sites), so that the it satisfies the condition
\be
\int d \xh \, \Omega(\xh) = 0.
\label{Omega-constraint}
\ee
Both in the liquid phase and in the gel phase, the system is uniform in density.  Therefore the saddle-point order parameter satisfie:
\be
\int \prod_{\beta ( \neq \alpha )} d \xv^{\beta}
\Omega(\xh) =
\left[ \overline{\langle \rho(\xv^{\alpha})
\rangle_{\rm m} }\right]_{\rm p} - \frac{N}{V} = 0.
\label{Omega-1-replica}
\ee
We therefore only have to consider the subspace of $\Omega(\xh)$ in which Eq.~(\ref{Omega-1-replica}) is obeyed. This subspace is usually called the {\em higher replica sector}.

The form of Landau free energy given in Eq.~(\ref{H_eff-def}) is generic for all isotropic networks:
\be
H_{\rm HR} =\int d\xh \left\{
 \frac{1}{2} \sum_{\alpha = 0}^n \left(
\nabla^{\alpha}  \Omega\right)^2
+ \frac{1}{2} r \, \Omega^2
- \frac{1}{3} w \, \Omega^3
+ \cdots \right\}.
\label{H_HR}
\ee
The coefficients in the Taylor expansion do of course depend on the short-scale structure of the constituent particles.   It may appear surprising that the cubic term in Eq.~(\ref{H_HR}) has a negative coefficient, which seems to suggest that the Landau theory is unstable, at lest for uniform values of the order parameter.  By Eqs.~(\ref{Omega^0},\ref{Omega^alpha}), however, a large and uniform $\Omega(\xh)$ also implies the large deviation of the particle densities from their equilibrium values in each replica [and therefore would violate Eq.~(\ref{Omega-1-replica})], and this would incur a large free-energy penalty.  The Landau theory of vulcanization is therefore locally stable.  Minimization of Eq.~(\ref{H_HR}) leads to the following saddle-point equation:
\be
- \sum_{\alpha = 0}^n\left(\nabla^{\alpha}\right)^{2} \Omega(\xv)
+ r \, \Omega(\xh) - w \, \Omega^2(\xh) = 0.
\label{Saddle-Eq}
\ee
This equation should be solved subject to the constraint~(\ref{Omega-1-replica}), and is therefore necessarily nonuniform in replicated space.

\subsection{Order parameter at the saddle-point}
By solving the saddle-point equation~(\ref{Saddle-Eq}), subject to the constraint~(\ref{Omega-1-replica}),  we obtain the saddle-point value of the order parameter, which we denote by $\overline{\Omega}(\xh)$.  For $r >0$, the saddle-point value of $\overline{\Omega} $ is trivially zero, corresponding to the liquid state.  For $r < 0$, the trivial saddle-point becomes unstable, and a nontrivial (i.e., replica-space dependent) saddle-point value emerges, corresponding to the amorphous solid state. It has the following form of superposed Gaussians~\footnote{Here, the bar indicates a saddle-point value, not a disorder average.}:
\be
\overline{\Omega}(\xh)
= q\, \int d\zv \int d \tau \, p(\tau) \,
\left({2 \pi}{\tau}\right)^{(1+n)d/2}
 e^{- \frac{\tau}{2}\sum_{\alpha = 0}^n(\xv^{\alpha} - \zv)^2}
 - \frac{\,\,q }{\quad V^{n}}.
 \label{Omega-bar}
\ee
The coefficient $q$ is identified with the number density of the infinite cluster (i.e., the gel fraction).  The dummy parameter $\tau$ is the inverse variance of the Gaussian fluctuations from their equilibrium positions of the localized particles (belonging to the infinite cluster), and is called the {\it inverse square localization length}.  The fact that there is a distribution of $\tau$ signifies the heterogeneous nature of randomly crosslinked systems: the infinite cluster is more rigid in some places and looser in others.  Both $q$ and $p(\tau)$ are determined by solving the saddle-point equation.  For details, see Refs.~\cite{XPMGZ-NE-VT,Vulcan_Goldbart}.  It is important to note that they are independent of strain deformation and all other parameters that are tunable in the measurement state.
The physics of the saddle-point order parameter can be better understood by focusing on the integrand inside Eq.~(\ref{Omega-bar}): The vector $\zv$ can be interpreted as the mean position of a particle in the gel fraction, and is localized around $\xv^0$, the position of the particle at the moment of crosslinking.  After crosslinking, the particle in the gel fraction ($\xv^{\alpha}, \alpha = 1, \ldots, n$) is localized around $\zv$, its mean location in the measurement state.  The uniform integral over $\zv$ implies that the gel is statistically homogeneous, after averaging over the quenched disorder.


If the crosslinker density is not high enough, the system is in the sol phase in the measurement state, no infinite cluster emerges, and all particles are delocalized, so that there is no correlation between their positions in different measurements, performed in the preparation and measurement states.  Hence, the associated joint pdf is a trivial constant, and the average order parameter is vanishes identically.   By contrast, in the gel phase,  there is an infinite cluster which constitutes a finite percentage of the overall mass.  For any particle in this infinite cluster, different measurements of its position are necessarily correlated: knowledge of its location in the preparation state tells us information of its whereabout in the measurement state, and hence the average order parameter in Eq.~(\ref{Omega-def}) is nonzero.  The order parameter $\Omega(\xh)$ thus distinguishes the gel phase from the sol phase.

We can actually deduce the functional form of the saddle-point order parameter~(\ref{Omega-bar}) using the definition of the order parameter~(\ref{Omega-def-physical}) [after subtracting a trivial constant, so that the constraint~(\ref{Omega-constraint}) is satisfied] together with some simple, intuitive reasoning.  We have already noted that particles belonging to the liquid fraction do not contribute to the saddle-point order parameter.  For a particle $i$ belonging to the infinite cluster, we expected that its position $\cv_i$ in the measurement state is Gaussian-distributed around its mean value, which we denote by $\zv_i$, with variance $1/\tau_i$.  The thermal averages inside Eq.~(\ref{Omega-def-physical}) combine to give
\ba
 \langle
\delta_{b^{\prime}}(\xv^1- \cv_i)
\rangle_{\rm m}
\cdots
\langle
\delta_{b^{\prime}}(\xv^n- \cv_i)
\rangle_{\rm m}
=
(2 \pi \tau_i)^{nd/2}\, \exp \left\{- \frac{\tau_i}{2}
\sum_{\alpha = 1}^n
|\xv^{\alpha} - \zv_i |^2\right\} .
\ea
The parameters $\zv_i$ and $\tau_i$ of course depend on the microscopic configuration of the preparation state at the instant of crosslinking, as well as on the crosslinks that are realized.  It is important to note that the mean position $\zv_i$  is distinct from the position of the particle at the instant of crosslinking, which we denote by $\cv_i^0$.   Although, for a given configuration of preparation state and crosslinker realization, $\zv_i$ is fully determined, it becomes a statistical variable after we include the various crosslinker realizations.  In fact, we expect that $\zv_i$ is Gaussian-distributed around the particle position $\cv_i^0$ in the preparation state, with the same variance $1/\tau_i$.   Also, for different realizations of crosslinkers, the variance $1/\tau_i$ itself should also be different, obeying some distribution $p(\tau)$.  An ergodic hypothesis suggests that this distribution is reflected in the microscopic spatial heterogeneity that individual samples of elastomer exhibit.  All these considerations suggest that
\ba
\overline{ \langle
\delta_{b^{\prime}}(\xv^1- \cv_i)
\rangle_{\rm m}
\cdots
\langle
\delta_{b^{\prime}}(\xv^n- \cv_i)
\rangle_{\rm m}  }
= \int d \zv\,
  \int d\tau\,
(2 \pi \tau)^{(1+n)\,d/2}\,p(\tau)\,
 \exp \left\{ - \frac{\tau}{2}
\sum_{\alpha = 1}^n
| \xv^{\alpha} - \zv|^2
- \frac{\tau}{2}
|\cv_i^0- \zv|^2
\right\} .
\label{average1-2-0}
\ea
Finally, we note that in the preparation state the system is a homogeneous liquid and thus translation invariant.  Hence, the position of a given particle $\cv_0$ is uniformly distributed over space, and therefore for an arbitrary function $f(\cv^0_i)$ we have
\be
\left[ f(\cv_i^0) \right]_{\rm p} = 
\int \frac{d\cv}{V}\,f(\cv).
\ee
We now substitute Eq.~(\ref{average1-2-0}) back into Eq.~(\ref{Omega-def-physical}) and carry out the average over the preparation state and also sum over all of the particles.  Thus, we finally arrive at the precise form of the saddle-point order parameter given in Eq.~(\ref{Omega-bar}).

\subsection{Translational symmetry and order parameter}

The $1+n$ replicated Hamiltonian $H_{1+n}$, Eq.~(\ref{H_1+n}), possesses the symmetry of independent translations of the replicas, i.e., it is invariant under the $\alpha$-dependent translations:
\be
\cv_i^{\alpha} \rightarrow
\cv_i^{\alpha} + {\bf u}^{\alpha},
\quad i = 1, \dots, N.
\nonumber
\ee
This symmetry is also possessed by the average order parameter $\overline{\Omega}(\xh)$ in the liquid phase (which has the value zero).  In the gel phase, however, this translational symmetry is explicitly broken by the saddle-point order parameter, Eq.~(\ref{Omega-bar}).  Translations of all particles in one replica leads to a distinct but energetically degenerate order parameter:
\be
\overline{\Omega}(\xv^0, \ldots, \xv^{\alpha}
+ \uv, \ldots \xv^n) \neq
\overline{\Omega}(\xv^0, \ldots,
\xv^{\alpha} , \ldots \xv^n) .
\nonumber
\ee
This reduction in in symmetry is a consequence of the localization of the particles associated with the emergence of an infinite cluster.  Nevertheless, translations of all replicas by a common vector $\uv$ remains a symmetry for Eq.~(\ref{Omega-bar}):
\be
 \overline{\Omega}(\xv^0,\xv^1, \ldots, \xv^n)  =
 \overline{\Omega}(\xv^0 + \bm{u},\xv^1
 + \bm{u}, \ldots, \cv^n + \bm{u}) .
\nonumber
\ee
This {\em macroscopic translational invariance} (MTI)~\cite{Vulcan_Goldbart} reflects the fact that gels and elastomers are {\em statistically\/} homogeneous.  This symmetry can be broken by translational ordering, such as smectic ordering in liquid crystalline elastomers.

\subsection{Classical theory of the elasticity of rubber}

Because rubbery materials can typically sustain large deformations, their elasticity theory is necessarily nonlinear.  This is already apparent in the classical theory of rubber elasticity, which gives the elastic free energy
\be
H(\bm{\Lambda}) =
\frac{\mu}{2} \Tr \bm{\Lambda}^{\rm T} \bm{\Lambda},
\label{eq:CTRExx}
\ee
in terms of the (spatially uniform) deformation gradient matrix $ \bm{\Lambda}$.
The matrix $\bm{\Lambda}$ relates the undeformed positions of particles
belonging to the infinite cluster to their deformed positions
$\zv' \equiv \bm{\Lambda} \cdot \zv$.
As most rubbery materials are nearly incompressible, ${\rm det}\, \Lambda$ can be taken to be unity. This is largely responsible for the nonlinear nature of rubber elasticity.  In the framework of statistical physics, the positions of particles fluctuate, and this deformation relation must be understood in the sense of relocation of the mean positions.
It is reasonable to suppose that the saddle-point order parameter of the deformed state is given by 
\be
\overline{\Omega}_{\bm{\Lambda}}(\xh)
= q\, \int_z \int_{\tau}
 e^{- \frac{\tau}{2}
  (\xv^0 - \zv)^2
 - \frac{\tau}{2}\sum_{\alpha = 1}^n
 |\xv^{\alpha} - \bm{\Lambda} \zv|^2}
 -  \frac{\,\,q}{\quad V^{n}}.
 \label{Omega-bar-Lm}
\ee
Note that the mean position $\zv$ is affinely deformed in the measurement state (replicas $1, \ldots, n$), but not in the preparation state (replica $0$), and that $q$ and $p(\tau)$ are not deformed (i.e., do not depend  on $\Lambda$).   Deformation of $0$-th replica can always be ``gauged away'' via a coordinate transform of the dummy integration variable $\zv$, and hence has no physical significance. It is the {\it relative} deformation between the preparation and measurement ensembles that has physical significance.

It has been shown explicitly~\cite{XPMGZ-NE-VT} that Eq.~(\ref{Omega-bar-Lm}) indeed satisfies the saddle-point equation and the boundary conditions for a solid having a uniform deformation $\bm{\Lambda}$.  Furthermore, by inserting Eq.~(\ref{Omega-bar-Lm}) into the Landau free energy~(\ref{H_HR})~\footnote{In particular, we need to follow the prescription in Eq.~(\ref{F-average-3}), expand the replicated free energy in powers of $n$, and the take the part linear in $n$. } it is straightforward to obtain the elastic free energy of the deformed system, as given in Eq.~(\ref{eq:CTRExx}), i.e., the classical theory of rubber elasticity~\cite{EL:Treloar}.  Moreover, one finds that $\mu \propto |r|^3$ for the shear modulus~\footnote{This $\mu$ (the shear modulus) is different from $\mu$ used in Sec.~\ref{sec:Deam-Edwards}, which there characterizes crosslinking density. }.  It is rather interesting to see that classical rubber elasticity theory emerges at the saddle-point level of vulcanization theory.

\subsection{Nematic elastomers and neo-classical elasticity theory}

In nematic elastomers, the polymer chains carry liquid crystalline mesogens that are prone to undergoing nematic ordering.  The mesogens can either be parts of the polymer backbone (i.e., main-chain nematic polymers) or attached to the backbone side-on (i.e., side-chain nematic polymer).  A Landau theory of nematic elastomers has be derived via vulcanization theory elsewhere~\cite{XPMGZ-NE-VT}; here we simply quote the results.

Let the (spatially uniform) nematic order parameter be $\Qm^0$ in the preparation state and $\Qm$ in the measurement state.
%
The free energy~(\ref{H_HR}) needs to be modified to include possible couplings between the nematic  and vulcanization order parameters.  To lowest order in these order parameters, the couplings are
\be
\int d \xh \left\{
\Qm^0_{ij} \,\,
\nabla_i^0\Omega \,\,
\nabla_j^0\Omega
+\sum_{\alpha=1}^{n}
\Qm_{ij}\,\,
\nabla_i^{\alpha}\Omega
\,\, \nabla_j^{\alpha}\Omega
\right\} .
\label{Q-Omega-Omega}
\ee
These couplings can be incorporated into the isotropic term in Eq.~(\ref{H_HR}) to yield
\be
H_{\mathrm HR} =\int d\xh
\left\{
 \frac{1}{2}
\bm{l}^0_{ij}\,
\nabla^{0}_i  \Omega\,
\nabla^{0}_j  \Omega
+ \frac{1}{2} \sum_{\alpha = 1}^n
\bm{l}_{ij}\,
\nabla^{\alpha}_i  \Omega\,
\nabla^{\alpha}_j  \Omega
+ \frac{1}{2} r \, \Omega^2
- \frac{1}{3} w \, \Omega^3
+ \cdots \right\},
\label{H_HR-Q}
\ee
where
\be
\bm{l}^0 \equiv \bm{I} + \Qm^0\quad{\mathrm and} \quad
\bm{l} \equiv \bm{I} + \Qm
\ee
are the polymer {\it step-length tensors\/} in the preparation and measurement states, respectively.

The general saddle-point in the presence of a macroscopic deformation gradient $\bm{\Lambda}$, which we denote by $\overline{\Omega}_{\bm{\Lambda}}(\xh)$, can also be obtained, by minimizing the free energy~(\ref{H_HR-Q}):
\be
\overline{\Omega}_{\bm{\Lambda}}(\xh)
= q\, \int_z \int_{\tau}
\exp\left\{- \frac{\tau}{2}
  \yv^0 \cdot \bm{l}^0 \cdot \yv^0
 - \frac{\tau}{2}\sum_{\alpha = 1}^n
 \yv_{\bm{\Lambda}}^{\alpha} \cdot \bm{l}
 \cdot \yv_{\bm{\Lambda}}^{\alpha}\right\}
- \frac{\,\,q}{\quad V^{n}},
 \label{Omega-bar-Lm-Q}
\ee
where
\be
\yv^0 \equiv \xv^0 - \zv
\quad{\rm and}\quad
\yv^{\alpha}_{\bm{\Lambda}}
\equiv \xv^{\alpha} - \bm{\Lambda} \cdot \zv.
\ee
Note that the Gaussian fluctuations of the particle positions have variance matrices $\bm{l}^0$ in the preparation state and $\bm{l}$ in the measurement state.  The effect of nematic order is therefore to render the thermal position fluctuations of each molecule anisotropic.

Substituting this deformed saddle-point~(\ref{Omega-bar-Lm-Q}) back into Eq.~(\ref{H_HR-Q}), and carrying out the prescription specified in Eq.~(\ref{F-average-3}), we find the following elastic free energy for nematic elastomers:
\be
H(\bm{\Lambda}) =
\frac{\mu}{2}\,\Tr\,\bm{l}^0 \bm{\Lambda}^{\rm T} \bm{l}^{-1} \bm{\Lambda},
\ee
which is precisely the {\it neo-classical elasticity theory\/} for nematic elastomers, derived by Warner and Terentjev~\cite{LCE:WT,WarTer96} as a generalization of the classical theory of rubber elasticity.

\section{Concluding remarks: Why vulcanization theory?}
\label{sec:conclusion}

It is quite satisfactory that the classical theory of rubber elasticity and the neo-classical elasticity theory of nematic elastomers can be derived from the vulcanization theory as saddle-point approximations. These theories were originally established by studying the statistics of single polymers inside a network, with two key assumptions~\cite{EL:Treloar,LCE:WT}: (1)~that strain deformations are affine at all scales; and (2)~that the polymer statistics are Gaussian.  These theories have attained classical status because of their remarkable successes in explaining the basic properties of isotropic and nematic elastomers.  Nevertheless, the starting point of single polymer statistics substantially restricts the generalizability of these theories.  For example, it is hard to see how one might systematically explore the nature of heterogeneities in random polymer networks using these molecular-level approaches.

Vulcanization theory was developed to understand the heterogeneities of randomly crosslinked systems.  In the face of its (perhaps heavy!) machinery---such as field theory, the replica technique, and multiple statistical ensembles---if all that could be obtained from it were re-derivations of classical results then it would be hard to argue that vulcanization theory has been worth investing in.  Fortunately, however, several fundamental new results have already been obtained using vulcanization theory, concerning the heterogeneity of rubbery materials: (1)~the distribution of localization lengths $p(\tau)$ in isotropic networks~\cite{0295-5075-28-7-011}; (2)~the distribution of internal random stresses in isotropic networks~\cite{Mao-07,PhysRevE.80.031140}; and (3)~the statistics of random fields and memory effects in isotropic-genesis nematic elastomers~\cite{Lu:2012kx}.  We  refer the reader to the literature for further details about these topics.  We think it is fair to say that we are still in an early stage in the unraveling of the statistical physics of network heterogeneities and memory effects in randomly crosslinked materials.  We hope that fresh and interesting results will be available to report in the near future.


We thank our numerous co-workers and several constructive anonymous referees for valuable collaborations and guidance. This work was supported by NSFC (China) via grant nos.~11174196 and 91130012, by the U.S.~NSF via DMR 09-06780 and DMR 12 07026, and by the Institute for Complex Adaptive Matter.


\end{document}